\documentclass[prl,twocolumn,superscriptaddress,showpacs]{revtex4}

\usepackage[dvips]{graphicx}
\usepackage{color}
\usepackage{latexsym}
\usepackage{amsmath}
\usepackage{amssymb}

\begin{document}

\preprint{LA-UR-08-1278}

\title{Combined experimental and theoretical investigation of the premartensitic transition in Ni$_2$MnGa}

\author{C.P. Opeil}
\affiliation{Physics Department Boston College, 140 Commonwealth
Avenue, Chestnut Hill, MA 02467}

\author{B. Mihaila}
\affiliation{Materials Science and Technology Division, Los Alamos
National Laboratory, Los Alamos, NM 87545}

\author{R.K. Schulze}
\affiliation{Materials Science and Technology Division, Los Alamos
National Laboratory, Los Alamos, NM 87545}

\author{L. Ma\~nosa}
\affiliation{Departament d\'{}Estructura i Constituents de la
Mat\`{e}ria, Facultat de F\'{i}sica, Universitat de Barcelona,
Diagonal 647, 08028 Barcelona, Catalonia, Spain}

\author{A. Planes}
\affiliation{Departament d\'{}Estructura i Constituents de la
Mat\`{e}ria, Facultat de F\'{i}sica, Universitat de Barcelona,
Diagonal 647, 08028 Barcelona, Catalonia, Spain}

\author{W.L. Hults}
\affiliation{Materials Science and Technology Division, Los Alamos
National Laboratory, Los Alamos, NM 87545}

\author{R.A. Fisher}
\affiliation{Materials Science and Technology Division, Los Alamos
National Laboratory, Los Alamos, NM 87545}

\author{P.S. Riseborough}
\affiliation{Physics Department, Temple University, Philadelphia, PA 19122}

\author{P.B. Littlewood}
\affiliation{Cavendish Laboratory, Madingley Road, Cambridge CB3
0HE, United Kingdom}

\author{J.L. Smith}
\affiliation{Materials Science and Technology Division, Los Alamos
National Laboratory, Los Alamos, NM 87545}

\author{J.C.~Lashley}
\affiliation{Materials Science and Technology Division, Los Alamos
National Laboratory, Los Alamos, NM 87545}



\begin{abstract}
Ultraviolet-photoemission (UPS) measurements and supporting specific-heat, thermal-expansion, resistivity and magnetic-moment measurements are reported for the magnetic shape-memory alloy Ni$_2$MnGa over the temperature range $100~K < T < 250~K$. All measurements detect clear signatures of the premartensitic transition ($T_\mathrm{PM}\sim 247~K$) and the martensitic transition ($T_\mathrm{M} \sim 196~K$). Temperature-dependent UPS shows a dramatic depletion of states (pseudogap) at $T_\mathrm{PM}$ located 0.3~eV below the Fermi energy. First-principles electronic structure calculations show that the peak observed at 0.3~eV in the UPS spectra for $T > T_\mathrm{PM}$ is due to the Ni-d minority-spin electrons. Below $T_\mathrm{M}$ this peak disappears, resulting in an enhanced density of states at energies around 0.8~eV. This enhancement reflects Ni-d and Mn-d electronic contributions to the majority-spin density of states and is accompanied by significant reconstruction of the Fermi surface.
\end{abstract}

\pacs{
81.30.Kf,         
71.20.Be,         
79.60.-i          
}

\maketitle

Ni-Mn-Ga alloys with near-stoichiometric compositions, Ni$_2$MnGa, are important functional materials~\cite{Magnetism05} owing to their magnetic shape-memory~\cite{Kakeshita02}, magnetocaloric~\cite{Marcos03} and magnetoresistive~\cite{Biswas2005} properties. This ferromagnetic fcc $L2_1$ Heusler ($a_\mathrm{fcc}=5.81$~\AA) was first identified by Webster \emph{et al.}~\cite{Webster1} as a system undergoing a martensitic transition (MT) in its ferromagnetic phase ($T_\mathrm{C}\sim380~K$) with little magnetic hysteresis. In the last decade research on these alloys has focused on the structural and magnetic characterization and on their shape-memory applications~\cite{Soderberg06}. First-principles calculations~\cite{Rabe, Lee02} and measurements on shape-memory alloys~\cite{Trevor, Ross, Dugdale, Lashley} indicate the driving role of the electronic structure and its relation to the lattice dynamics.

\begin{figure}[t]
   \includegraphics[width=\columnwidth]{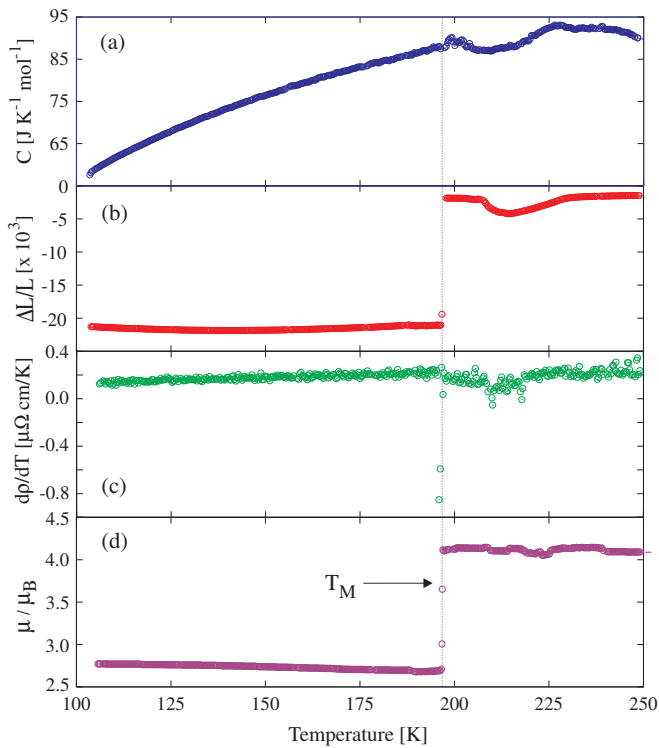}
   \caption{\label{fig1} (Color online)
   Temperature dependence of the heat capacity~(a), thermal expansion~(b), temperature-derivative of the resistivity~(c), and magnetic moment~(d) in the region of the PMT in Ni$_2$MnGa.}
\end{figure}

The lattice dynamics of Ni$_2$MnGa has been investigated from ultrasonic measurements~\cite{Worgull96} and neutron diffraction experiments~\cite{Zheludev95,manosa01,Steve07}. It was found that the transverse TA$_2$ phonon branch exhibits pronounced softening at $1/3$ of the zone boundary on decreasing the temperature, and this softening was described as a Bain distortion in the context of the Wechler, Lieberman, and Read theory of martensite formation~\cite{Lieberman}. In similar structural shape-memory alloys InTl~\cite{Trevor}, AuZn~\cite{Ross} and NiAl~\cite{Dugdale}, this softening is associated with nesting features of the Fermi surface~\cite{Lee02}. Below a certain temperature, there is a freezing of the displacements associated with this soft phonon so that a micro-modulated phase forms, which is described as a periodic distortion of the parent cubic phase~\cite{Zheludev95}. In Ni$_2$MnGa, the premartensitic phase develops with little or no thermal hysteresis and is driven by a magnetoelastic coupling~\cite{Planes97}. On further cooling, Ni$_2$MnGa transforms to an approximately five-layered quasi-tetragonal martensitic structure. The low-temperature phase is \emph{incommensurate}~\cite{Zheludev95} with a period (0.43,0.43,0) and exhibits well-defined phasons best characterized as charge-density wave (CDW) excitations~\cite{Steve07}.

In this paper we study the role of conduction electrons in the two-step MT in Ni$_2$MnGa using photoemission spectroscopy and thermodynamic measurements. LEED and X-ray diffraction Laue measurements show the quality of our sample is appropriate for high-resolution photoemission spectroscopy. Ultraviolet photoemission (UPS) measurements show the opening of a pseudogap at 0.3 eV below the Fermi energy at the MT and provide further evidence that the Fermi surface is strongly nested at the MT and is only partially nested at the premartensitic transition (PMT).

Single crystals were grown by a Bridgman technique. All samples were spark cut from a large Ni$_2$MnGa single crystal and used for thermal expansion, specific heat, resistivity, magnetic moment, X-ray diffraction and UPS experiments. We note that in the low-temperature phase, the crystal structure of our sample is found to be a five-layered martensite (termed $5R$ or $10M$): the unit cell is monoclinic with the parameters $a$=4.2\AA, $b$=5.5\AA, $c$=21\AA, $\alpha$=$\gamma$=90$^\circ$ and $\beta$=91$^\circ$~\cite{manosa01}.  The linear coefficient of thermal expansion was measured in a three-terminal capacitive dilatometer~\cite{George} over the range of 100~K $<$ T $<$ 250~K. The specific heat was measured using a thermal-relaxation method~\cite{Lashley1}. The magnetic moment was measured using a vibrating-sample magnetometer on a Quantum Design Physical Properties Measurement System (PPMS). Resistivity measurements were made on the PPMS using an ac technique.

Figure~\ref{fig1} shows clear signatures of a PMT and MT in the specific heat, thermal expansion, temperature-derivative of the resistivity and the magnetic moment. The specific-heat data show a broad peak centered at $T=225~K$ in agreement with previous measurements in a sample of the same composition~\cite{hc}. This feature does not have any measurable thermal hysteresis associated with it. Conversely, the MT at $T_\mathrm{M}=196~K$ is a sharp, first-order transition associated with 8~K of thermal hysteresis. Low-temperature specific-heat measurements (not shown) give an electronic specific heat coefficient, $\gamma = 10.6$~mJ~K$^{-2}$mol$^{-1}$, and Debye temperature, $\Theta_D = 205.6$~K. The high-temperature effects are mirrored in the thermal expansion where the sharp, discontinuous MT is preceded by a broad feature with an onset temperature of $T_\mathrm{PM}=247~K$. The temperature-derivative of the resistivity shows a break in the resistivity slope at $T=214~K$ followed by a discontinuity at $T_\mathrm{M}=196~K$, as does the magnetic moment. The lack of thermal hysteresis at the PMT and the behavior of the thermal expansion, specific heat and electrical resistivity is consistent with CDW formation: a continuous transition associated with CDW onset (PMT), and discontinuous behavior at lock-in (MT)\cite{cravenmeyer77}.

\begin{figure}[t]
   \includegraphics[width=\columnwidth]{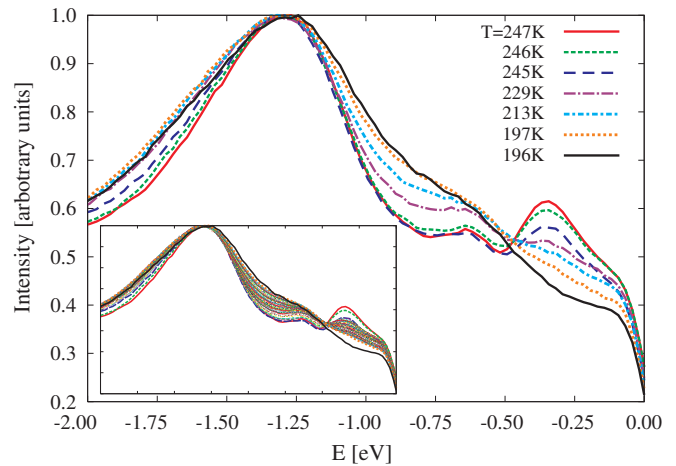}
   \caption{\label{fig2} (Color online)
   Energy dependence of the normal photoemission intensity for selected temperatures in the region of the PMT in Ni$_2$MnGa. The inset shows \emph{all} intensity plots at 1~K temperature intervals.}
\end{figure}

Further insight into the nature of the PMT can be obtained using temperature-dependent photoemission spectroscopy~\cite{photo-exp}. Figure~\ref{fig2} illustrates the normal-emission UPS spectra (He~I, $h\nu$=21.2~eV) for several temperatures of interest. In addition, the inset in Fig.~\ref{fig2} shows the UPS spectra in the temperature range of the PMT at~1K temperature intervals. In the inset, two sudden redistributions of the UPS intensity are observed corresponding to the onset of the PMT and the MT, respectively. Above $T_\mathrm{PM}$ and below $T_\mathrm{M}$ the UPS spectra remain unchanged and are not shown. For $T > T_\mathrm{PM}$, the UPS spectrum exhibits a prominent peak located at a binding energy of 1.3~eV, together with two smaller features at 0.8~eV and 0.3~eV. As the temperature is lowered below $T_\mathrm{PM}$, we note a rapid drop in the intensity of the 0.3~eV peak, followed by a slow, but continuous, decrease in the intensity of this peak with decreasing temperature. A second sudden drop in the intensity of this peak occurs just above $T_\mathrm{M}$. Throughout the process of depletion of the allowed electronic states close to the Fermi energy, we notice an enhancement in the number of available states above 0.8~eV. The temperature dependence of the UPS spectra reported here indicates the formation of a shallow pseudogap in the allowed density of states, in the PMT region, for energies between 0.3~eV and the Fermi energy. Similar effects have been reported in the past for high-$T_\mathrm{C}$ cuprates~\cite{cuprates} and purple bronze~\cite{purple}.

We note that the UPS spectra depicted in the inset of Fig.~\ref{fig2} show a much sharper transition at $T_\mathrm{M}$ then $T_\mathrm{PM}$, consistent with the data illustrated in Fig.~\ref{fig1}. The thermal expansion, resistivity and magnetic moment data support the notion of a first-order transition at the MT, $T_\mathrm{M}$. The smooth nature of the PMT makes difficult the identification of the onset temperature for the PMT, $T_\mathrm{PM}$. This is perhaps best illustrated by the thermal expansion data.  Nevertheless, from the UPS and thermal expansion data, we estimate $T_\mathrm{M} - T_\mathrm{PM} \approx $~50~K.

\begin{figure}[t]
   \includegraphics[width=\columnwidth]{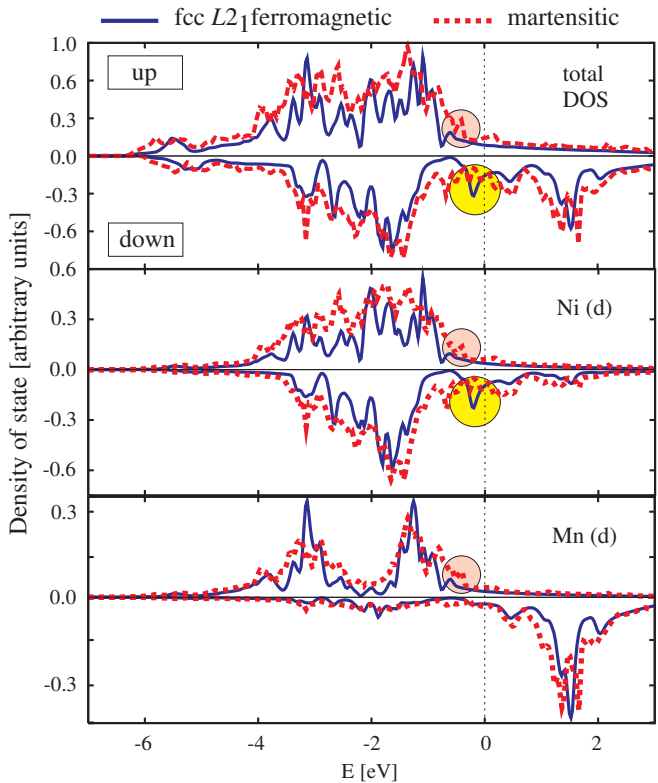}
   \caption{\label{fig3} (Color online)
   The majority- and minority-spin DOS for the spin-polarized martensitic phase ($T < T_\mathrm{M}$) and the spin-polarized fcc ($T_\mathrm{PM} < T < T_\mathrm{C}$) of  Ni$_2$MnGa. The top panel depicts the total DOS, whereas the middle and bottom panels depict the contributions due to the d~electrons of the Ni and Mn atoms, respectively. The contributions due to the Ga electronic degrees of freedom are smaller by an order of magnitude and have been disregarded for the purpose of this comparison.}
\end{figure}

Our UPS data was compared with results of first-principles band-structure calculations using the generalized gradient approximation approach~\cite{GGA} in the full-potential linearized-augmented-plane-wave method~\cite{Blaha}. We use the experimental lattice constants and calculations are performed on grids of 286 and 726~$k$ points in the irreducible Brillouin zone for the austenitic and martensitic phases, respectively. Figure~\ref{fig3} depicts the majority-spin (spin-up) and minority-spin (spin-down) densities of states (DOS) for the spin-polarized martensitic phase ($T < T_\mathrm{M}$) and the spin-polarized fcc phase ($T_\mathrm{PM} < T < T_\mathrm{C}$) of bulk Ni$_2$MnGa. The main contributions to the total density of states in either phase or spin state are due to the Ni-d and Mn-d electrons. We note that our electronic structure results also show a redistribution of the density of states away from the Fermi surface in the martensitic phase as compared to the fcc phase. Our calculations indicate that the missing spectral weight, observed experimentally near the Fermi energy in the martensitic phase, is due to the disappearance of the Ni-d peak located at 0.3~eV in the spin-down DOS corresponding to the fcc phase. In the martensitic phase, the spectral weight shifts to slightly lower binding energies and results in the enhancement of the peak observed at 0.8~eV in the spin-up DOS. The latter peak has a combined Ni-d and Mn-d character, as reflected by the peaks highlighted in the plots of the Ni-d and Mn-d spin-up partial DOS. Although we predict that near the Fermi surface there is a spectral-weight transfer from minority- to spin-up electrons (which could potentially be resolved by spin-polarised photoemission), the net magnetization shown in Fig.~1d in fact drops at the transition. The calculated net magnetization change (integrated over all energies) is found to be close to zero, since the low energy redistribution of spin density is approximately canceled by higher energy states. The disagreement with experiment indicates that correlation effects may be important for the deeper Mn d-levels - note that the magnetization is mostly dominated by Mn whereas the Fermi surface is preponderantly from Ni states.

\begin{figure}[t]
   \includegraphics[width=\columnwidth]{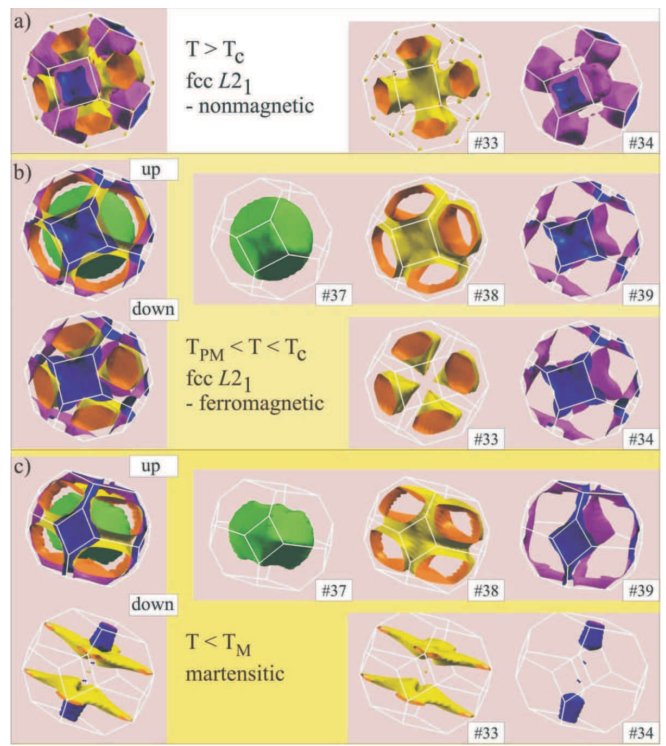}
   \caption{\label{fig4}
   Fermi surfaces in the Brillouin zones corresponding to a)~the non-magnetic fcc phase ($T > T_\mathrm{C}$), b)~the spin-polarized fcc phase ($T_\mathrm{PM} < T < T_\mathrm{C}$), and c)~the martensitic ($T < T_\mathrm{PM}$) of Ni$_2$MnGa. Both the merged and the individual band contributions to the Fermi surface are depicted.
   Plots performed with XCrySDen~\cite{xcrysden} using the
   structural data from Ref.~\cite{struct_data}.}
\end{figure}

The Fermi surfaces (FS) in the Brillouin zones corresponding to the various phases of Ni$_2$MnGa are displayed in Fig.~\ref{fig4}. Two electronic bands contribute to the FS in the non-magnetic phase ($T > T_\mathrm{C}$), whereas three and two electronic bands contribute to the spin-up and spin-down FS below $T_\mathrm{C}$. The character of the spin-down Fermi surfaces for $T_\mathrm{PM} < T < T_\mathrm{C}$ is dominated by the Ni-d electrons, whereas the character of the spin-up Fermi surfaces is shared by both Ni-d and Mn-d electrons. Below $T_\mathrm{M}$ the spin-up contribution to the FS look similar to the spin-up contribution above $T_\mathrm{M}$, but the FS is nested along the $k_z$ direction in the martensitic phase. The change in the FS of the spin-down electrons above $T_\mathrm{PM}$ and below $T_\mathrm{M}$ is more dramatic and reflects the changes noted in the calculated DOS and the measured UPS spectra. We note that the PMT is not accompanied by a change in the magnetic moment, whereas $T_M$ marks a substantial rearrangement of the spin density {\em between} bands, as seen in both experiment and theory. Thus we speculate that the transition at $T_{PM}$ corresponds to a nesting feature of a single band. As the amplitude of this instability grows on lowering the temperature, mixing with other bands becomes inevitable - which apparently triggers a second and stronger instability in the martensitic phase involving a redistribution between the spin directions and the development of quasi-1D-like reconstructed Fermi surface for the two spin-down bands.

To summarize, in this paper we report the presence of a pseudogap in photoemission spectra, 0.3~eV below the Fermi energy, at the PMT temperature, $T_\mathrm{PM}$. Based on first-principles electronic structure calculations, we conclude that the changes in the experimental photoemission spectra are due to a redistribution of the electronic density of states associated with the Mn-d and Ni-d electronic degrees of freedom. We show that the peak observed at 0.3~eV in the UPS spectra above $T_\mathrm{PM}$ is due to the Ni-d spin-down electrons. Below $T_\mathrm{M}$ this peak disappears, resulting in an enhanced DOS at energies around 0.8~eV. This enhancement reflects Ni-d and Mn-d electronic contributions to the spin-up DOS. The calculated innermost Fermi surface of spin-up electronic states exhibits weak momentum dispersion, which, together with the changes in the material properties observed experimentally in the premartensitic temperature region (see Fig.~\ref{fig1}), suggests a CDW pseudogap-opening mechanism from Fermi surface nesting in Ni$_2$MnGa.

\begin{acknowledgments}
This work was supported in part by the Los Alamos National Laboratory under the auspices of the U.S. Department of Energy, and the Trustees of Boston College. LM and AP acknowledge partial support from projects MAT2007-61200 (CICyT, Spain) and  2005SGR00969 (DURSI, Catalonia). PSR acknowledges support from the U.S. Department of Energy through award DEFG02-01ER45872. The authors would like to acknowledge useful conversations with Y.~Lee and B.~Harmon.
\end{acknowledgments}


\begin{thebibliography}{99}

\bibitem{Magnetism05}
{\it Magnetism and Structure in functional materials}
ed. A. Planes, L. Ma\~nosa, and A. Saxena,
Materials Science Series, Vol. 79 (Springer Verlag, Berlin 2005).


\bibitem{Kakeshita02}
T. Kakeshita and K. Ullakko,
MRS Bull. \textbf{27}, 105 (2002).

\bibitem{Marcos03}
J. Marcos \emph{et al.},
Phys. Rev. B \textbf{68}, 094401 (2003).

\bibitem{Biswas2005}
C. Biswas, R. Rawat and S.R. Barman,
Appl. Phys. Lett. \textbf{86}, 202508 (2005).

\bibitem{Webster1}
P.J. Webster \emph{et al.},
Philos. Mag. B \textbf{49}, 295 (1984).

\bibitem{Soderberg06}
See O. Soderberg \emph{et al.}, 
{\it Giant magnetostrictive materials}
in {\it Handbook of Magnetic Materials}, Vol. 79, ed. K.J.H. Buschow
(Elsevier, Amsterdam, 2006) and references therein.

\bibitem{Rabe}
C. Bungaro, K. M. Rabe, and A. Dai Corso,
Phys. Rev. B \textbf{68}, 134104 (2003);
A.T. Zayak \emph{et al.}, 
Phys. Rev. B \textbf{68}, 132402 (2003);
G.I. Velikohkhatnyi and I.I. Naumov,
Phys. Sol. State \textbf{41}, 617 (1999).

\bibitem{Lee02}
Y. Lee, J.Y. Rhee and B.N. Harmon,
Phys. Rev. B \textbf{66}, 054424 (2002);
P.~Entel \emph{et al.},
J. Phys. D: Appl. Phys. \textbf{39}, 865 (2006).

\bibitem{Trevor}
M. Liu, T. R. Finlayson, and T. F. Smith,
Phys. Rev. B {\bf48}, 3009 (1993).

\bibitem{Ross}
P.A. Goddard \emph{et al.},
Phys. Rev. Lett. \textbf{94}, 116401 (2005).

\bibitem{Dugdale}
S.B. Dugdale \emph{et al.},
Phys. Rev. Lett. \textbf{96}, 046406 (2006).

\bibitem{Lashley}
J.C. Lashley \emph{et al.},
Phys. Rev. B \textbf{75}, 205119 (2007);
G. Jakob, T. Eichhorn, M. Kallmayer, and H. J. Elmers,
Phys. Rev. B \textbf{76}, 174407 (2007).

\bibitem{Worgull96}
J. Worgull, E. Petti, and J. Trivisonno,
Phys. Rev. B \textbf{54}, 15695 (1996);
M. Stipcich \emph{et al.},
Phys. Rev. B \textbf{70}, 054115 (2004).

\bibitem{Zheludev95}
A. Zheludev \emph{et al.},
Phys. Rev. B \textbf{51}, 11310 (1995);
\textit{ibid.} Phys. Rev. B \textbf{54}, 15045 (1996)

\bibitem{manosa01}
L. Ma\~nosa \emph{et al.},
Phys. Rev. B \textbf{64}, 024305 (2001).

\bibitem{Steve07}
S.M. Shapiro \emph{et al.}, 
Euro. Phys. Lett. \textbf{77}, 56004 (2007).

\bibitem{Lieberman}
M.S. Wechlsher, D.S. Lieberman, and  T.A.~Read,
J. Metals \textbf{5}, 1503 (1953).

\bibitem{Planes97}
A. Planes, E. Obrad\'o, A. Gonz\`alez-Comas and L.~Ma\~nosa,
Phys. Rev. Lett. \textbf{79}, 3926 (1997).

\bibitem{George}
G.M. Schmeideshoff \emph{et al.},
Rev. Sci. Instrum. \textbf{77}, 123907 (2006).

\bibitem{Lashley1}
J.C. Lashley \emph{et al.},
Cryogenics \textbf{43}, 369 (2003).

\bibitem{hc}
   F.J. P\'erez-Reche, E. Vives, L. Ma\~nosa, and A. Planes,
   Mater. Sci. Eng. A \textbf{378}, 353  (2004).

\bibitem{cravenmeyer77}
R.A. Craven and S.F. Meyer, Phys. Rev. B \textbf{16}, 4583 (1977).

\bibitem{photo-exp}
   The crystal surface preparation and details of the experimental setup have been discussed in C.P. Opeil \emph{et al.}, Phys. Rev. B \textbf{73}, 165109 (2006); \emph{ibid.} \textbf{75}, 045120 (2006). The sample surface quality and alignment were verified using LEED measurements.

\bibitem{cuprates}
    M.R. Norman, D. Pines, and C. Kallin,
    Adv. Phys. \textbf{54}, 715 (2005).

\bibitem{purple}
   M.A. Valbuena \emph{et al.}, 
   J. Phys. Chem. Solids \textbf{67}, 213 (2006).

\bibitem{GGA}
J.P. Perdew, K. Burke, M. Ernzerhof,
Phys. Rev. Lett. \textbf{77}, 3865 (1996).

  \bibitem{Blaha}
  P. Blaha \emph{et al.}, 
  WIEN2k, An Augmented Plane Wave Plus Local Orbitals Program for Calculating
  Crystal Properties (Karlheinz Schwarz, Technische Universität Wien, Austria,
  2001).

\bibitem{xcrysden}
  A. Kokalj,
  Comp. Mater. Sci. \textbf{28}, 155 (2003).
  Code available from http://www.xcrysden.org/.

\bibitem{struct_data}
  D.Y. Cong, P. Zetterstr\"om, and Y.D. Wang,
  Appl. Phys. Lett. \textbf{87}, 111906 (2005).

\end{thebibliography}
\end{document}